\documentclass[twocolumn,showpacs,prl]{revtex4}% Physical Review B
\usepackage{bm}
\usepackage{epsfig}
\newcommand{\nix}[1]{}

\def\gsim{\ \raise 3pt \hbox{$>$} \kern -8.5pt \raise -2pt \hbox{$\sim$}\ }
\def\lsim{\ \raise 3pt \hbox{$<$} \kern -8.5pt \raise -2pt \hbox{$\sim$}\ }

\newcommand{\pe}{_{\rm pe}}
\newcommand{\Be}{_{\rm Be}}
\newcommand{\st}{_{\rm st}}

\begin{document}

\title{
Diffusive Radiation in One-dimensional Langmuir Turbulence }
\author{G.D.~Fleishman$^{1}$, I.N.~Toptygin$^2$}
\affiliation{$^1$A.F.~Ioffe Physico-Technical Institute, Russian
Academy of Sciences, 194021 St.~Petersburg, Russia; New Jersey
Institute of Technology, Newark, NJ 07102}
\affiliation{$^2$State Polytechnical University, 195251
St.~Petersburg, Russia}

%\date{\today}

\begin{abstract}
We calculate spectra of radiation produced by a relativistic
particle in the presence of one-dimensional Langmuir turbulence
which might be generated by a streaming instability in the plasma,
in particular, in the shock front or at the shock-shock
interactions. The shape of the radiation spectra is shown to
depend sensitively on the angle between the particle velocity and
electric field direction. The radiation spectrum in the case of
exactly transverse particle motion is degenerate and similar to
that of spatially uniform Langmuir oscillations. In case of
oblique propagation, the spectrum is more complex, it consists of
a number of power-law regions and may contain a distinct
high-frequency spectral peak. %at $\omega=2\omega\pe \gamma^2$.
The emission process considered is relevant to various laboratory
plasma settings and for astrophysical objects as gamma-ray bursts
and collimated jets.
\end{abstract}

\maketitle

%\section{Introduction}

Various kinds of two-stream instability including longitudinal,
transverse, and oblique modes
\cite{Buneman_1958,Weibel_1959,Fainberg_etal_1970,Bret_etal_2006}
are common for many astrophysical objects
\cite{Gallant_Arons_1994,Fender_Belloni_2004,Nishikawa_etal_2005}
and laboratory plasma settings
\cite{Tabak_etal_1994,Romanov_etal_2004}. These instabilities are
efficient to produce a high level of magnetic and/or electric
turbulence in the plasma. We note that the electric turbulence is
of primary importance for many applications, in particular, it is
capable of charged particle acceleration
\cite{Bykov_Uvarov_1993,Frederiksen+2004,Nishikawa_etal_2005} in
contrast to purely magnetic turbulence, which is only efficient
for angular scattering of the charged particle while cannot change
the particle energy directly. Remarkably, recent theoretical and
numerical studies \cite{Bret_etal_2005,Dieckmann_2005,Silva_2006}
demonstrate that the energy density going to the electric
(Langmuir) turbulence can be very large, exceeding, for example,
the energy density of the initial regular magnetic field in the
plasma.

Relativistic electrons propagating in the plasma with developed
Langmuir turbulence will random walk due to random Lorenz forces
produced by these turbulent electric fields. Apparently, during this
random diffusive motion the electron will generate electromagnetic
emission. We will refer to this radiative process as
\emph{Diffusive} Radiation in Langmuir waves (DRL) to emphasize the
key role of the diffusive random walk of the particle in the
stochastic electric fields. % like in case of the \emph{Diffusive}
%Synchrotron Radiation (DSR) in random magnetic fields
%\cite{Topt_Fl_1987,Topt_etal_1987,Fl_2005b,Fl_2005a}.

Curiously, the theory of DRL has not been developed in sufficient
detail yet, although a few particular issues related to this
radiative process (called also \emph{Electrostatic
Bremsstrahlung}) have been considered
\cite{Gailitis_Tsytovich_1964,
Colgate_1967,Tsytovich_Chikhachev_1969, Melrose_1971,
Kaplan_Tsytovich_1973, Schlickeiser_2003}. However, the detailed
treatment of the DRL spectral shape in various regimes is
currently unavailable. This letter attempts to remedy the
situation by calculating the DRL spectrum within the perturbation
theory and determines the region of applicability of the
perturbative treatment.

%\section{Perturbation theory of DRL}
%\label{S_Pert_DRL}

Electromagnetic emission produced by a charged particle can be
calculated within the perturbation theory when  the particle moves
almost rectilinearly with almost constant velocity. Apparently, the
non-zero acceleration of the particle in the external field should
be taken into account to obtain a non-zero radiation intensity. This
perturbative treatment is widely used because of its simplicity.
Frequently, one calculates first the particle acceleration ${\bf
w}(t)$ due to a given field along the rectilinear trajectory and
then uses this expression obtained for ${\bf w}(t)$ to find the
radiation spectrum. In case of a random external field, however,
when ${\bf w}(t)$ is also a random function of time $t$ it is more
convenient to express the radiation intensity via spatial and
temporal spectrum of the
external electric and/or magnetic field directly %.
%
%Within theoretical formulation presented in
\cite{Fl_2005a}. Accordingly, the spectral and angular distribution
of the emission produced by a
single particle %with the Lorenz-factor $\gamma$
in a plasma with random field has the form:
\begin{equation}
\label{cal_E_wFn_perp_2}
  W_{\Omega,\omega}^{\bot}=\frac{(2\pi)^3Q^2}{M^2c^3\gamma^2V}
  \left(\frac{\omega}{\omega
  '}\right)^2
  \left[1-\frac{\omega}{\omega' \gamma_*^2} + \frac{\omega^2}{2\omega'^2 \gamma_*^4}
\right] \times $$$$
    \int  dq_0 d{\bf q}
  \delta(\omega'-q_0+{\bf qv}) \mid
 {\bf F}_{q_0, {\bf q}\bot} \mid^2,
\end{equation}
where $\gamma_* =
\left(\gamma^{-2}+\frac{\omega\pe^2}{\omega^2}\right)^{-1/2}$, $Q$,
$M$, and $\gamma$ are the charge, mass, and Lorenz-factor of the
emitting particle, $c$ is the speed of light, $\omega\pe$ is the
plasma frequency, $V$ is the volume of the emission source, ${\bf
F}_{q_0, {\bf q}\bot}$ is the temporal and spatial Fourier component
of the Lorenz force transverse to the emitting particle velocity,
\begin{equation}
\label{om_prime_def}
  \omega'=\frac{\omega}{2}\left(\gamma^{-2}+\theta^2 + \frac{\omega\pe^2}{\omega^2}
  \right),
\end{equation}
$\theta$ is the emission angle relative to the particle velocity
vector ${\bf v}$, $\omega$ is the frequency of the emitted wave.
Contribution $W_{\Omega,\omega}^{\bot}$ (marked with the superscript
$\bot$) is provided by a component of the particle acceleration
transverse to the particle velocity. In case of the electric ${\bf
E}$ (in contrast to magnetic) field, there is also a component of
the acceleration along the particle velocity. The corresponding
contribution has the form
\begin{equation}
\label{cal_E_w_par_2}
  W_{\Omega,\omega}^{\|}=\frac{2(2\pi)^3Q^4}{M^2c^3\gamma^6V}  \left(\frac{\omega}{\omega
  '}\right)^3
  \left[1-\frac{\omega}{2\omega' \gamma_*^2}
\right]  \times $$$$
    \int  dq_0 d{\bf q}
  \delta(\omega'-q_0+{\bf qv}) \mid
 {\bf E}_{q_0, {\bf q}\|} \mid^2,
\end{equation}
which is typically small by a factor $\gamma^{-2}$ compared with
the transverse contribution. Nevertheless, there exist special
cases  when the transverse contribution is zero or very small and
the parallel contribution comes to play. In particular, for the
considered here one-dimensional turbulent electric field the
parallel contribution will dominate for a particle moving along
the field direction.

Let us start with the case when the particle moves at a large
angle to the Langmuir turbulence direction $\vartheta \gg
\gamma^{-1}$, so the standard transverse contribution dominates.
Following the derivation given in \cite{Fl_2005a}, but with the
Lorenz force ${\bf F} = Q {\bf E}$ specified by electric ${\bf E}$
in place of magnetic ${\bf B}$ field, it is easy to find
\begin{equation}
\label{E_corr}
     \mid {\bf F}_{q_0, {\bf q}\bot} \mid^2 = Q^2 \mid {\bf E}_{q_0,{\bf q}\bot} \mid^2
  = $$$$   \frac{TV}{(2\pi)^4} Q^2 \left(\delta_{\alpha\beta} -
\frac{v_{\alpha}v_{\beta}}{v^2}\right) K_{\alpha \beta}(q_0,{\bf
q}),
   %= \frac{TV}{(2\pi)^4}
%   K(q_0,{\bf q}) %\delta(q_0-q_0({\bf q}))
\end{equation}
where $K_{\alpha\beta}(q_0,{\bf q}) = C_{\alpha\beta} K(q_0,{\bf
q})$, $T$ is the total time of emission; $C_{\alpha\beta}$ describes
the longitudinal nature of the Langmuir waves, i.e.,
$C_{\alpha\beta}=
q_{\alpha}q_{\beta}/q^2$, %which reduces to
%$C_{\alpha\beta}=n_{\alpha}n_{\beta}$ for one-dimensional
%turbulence,
while $K(q_0,{\bf q})$ is the temporal and spatial
spectrum of the Langmuir turbulence.

The developed approach allows for arbitrary anisotropy of the
turbulence, although we have to specify the shape of the turbulence
spectrum to promote further the calculation of the DRL spectrum.
Analytical and numerical studies of the two-stream instabilities
suggest that the Langmuir turbulence produced is frequently highly
anisotropic \cite{Buneman_1958,Fainberg_etal_1970,Bret_etal_2006,
Bret_etal_2005,Dieckmann_2005,Silva_2006}, which is confirmed also
by available \textit{in situ} observations, e.g., performed in the
Earth magnetosphere \cite{Mangeney_etal_2006}. Although this
anisotropy can be reduced at later stages of the nonlinear
turbulence evolution due to randomization of the wave vector
directions, here we assume that the Langmuir turbulence is highly
anisotropic, namely, one-dimensional. As an example, we can suppose
that all the wave vectors are directed along the shock normal ${\bf
n}$, therefore, $C_{\alpha\beta}=n_{\alpha}n_{\beta}$, while the
spectrum $K(q_0,{\bf q})$ can be approximated by a power-law over
$q_\|$ above certain critical value $k_0$:
\begin{equation}
\label{E_tran_Lang_spectr}
     K(q_0,{\bf q})
   =  \frac{a_\nu k_0^{\nu-1} \left<E_L^2\right> }{(k_0^2+q_\|^2)^{\nu/2}}
   \delta({\bf q_\bot})
  \delta(q_0-\omega\pe).
\end{equation}
Here, the presence of the second $\delta$-function is related to
the assumption that the electric turbulence is composed of
Langmuir waves all of which oscillate in time with the same
frequency $\omega\pe$; the normalization constant $a_\nu$ is set
up by the condition $\int K(q_0,{\bf q}) dq_0 d{\bf q} =
\left<E_L^2\right>$, where  $\left<E_L^2\right>$ is the mean
square of the electric field in the Langmuir turbulence.

Now, substituting (\ref{E_corr}) with (\ref{E_tran_Lang_spectr})
into general expression (\ref{cal_E_wFn_perp_2}), taking the
integrals over $dq_0$, $dq_\|$, and $dq_\bot$ with the use of three
available $\delta$-functions and dividing by the total (infinite)
time of emission $T$ we find
\begin{equation}
\label{I_Lang_w_spectr}
  I_{\Omega,\omega}^{\bot}=\frac{a_\nu k_0^{\nu-1} \left<E_L^2\right>  Q^4 \sin^2\vartheta}
  {2\pi M^2c^4\gamma^2 |\cos\vartheta|}  \left(\frac{\omega}{\omega
  '}\right)^2 \times $$$$
  \left[1-\frac{\omega}{\omega'\gamma_*^{2}} +
  \frac{\omega^2}{2\omega'^2\gamma_*^{4}}
  \right] % \times $$$$ %\\\\
   \left\{\left(\frac{\omega' -\omega\pe}{\bf nv}\right)^2
   +k_0^2\right\}^{-\nu/2}
   ,
\end{equation}
where $\vartheta$ is the angle between the particle velocity ${\bf
v}$ and vector ${\bf n}$. This expression diverges formally when
$\cos\vartheta \rightarrow 0$. Accordingly, for a particle moving
transversely to Langmuir turbulence direction we need to
recalculate the integrals in Eq. (\ref{cal_E_wFn_perp_2}) taking
into account that $\delta(\omega'-q_0+{\bf qv}) \rightarrow
\delta(\omega'-q_0)$ for ${\bf qv}=0$, which yields
\begin{equation}
\label{I_Lang_nw_2}
%\begin{array}l
  I_{\Omega,\omega}^{\bot}=\frac{Q^4 \left<E_L^2\right>}
  {2\pi M^2c^3\gamma^2}  \left(\frac{\omega}{\omega
  '}\right)^2
  \left[1-\frac{\omega}{\omega'\gamma_*^{2}} +
  \frac{\omega^2}{2\omega'^2 \gamma_*^{4}}\right] %\\\\
   \delta(\omega'-\omega\pe)
%\end{array}
\end{equation}
in full agreement with the results of
\cite{Tsytovich_Chikhachev_1969} obtained for spatially uniform
Langmuir oscillations. This is not a random coincidence. Indeed,
since the 1D Langmuir waves experience spatial variations along only
one direction (of vector ${\bf n}$), thus, the particle moving
transversely to this direction "feels" spatially uniform field
pattern like that considered in \cite{Tsytovich_Chikhachev_1969}.

There is another special geometry when intensity
(\ref{I_Lang_w_spectr}) is insufficient to describe the DRL
spectrum: it is the case of particle motion along vector ${\bf
n}$. Indeed, the "parallel" contribution becomes important for
$\vartheta \lsim \gamma^{-1}$. Substituting
(\ref{E_tran_Lang_spectr}) into (\ref{cal_E_w_par_2}) and taking
the integrals similarly to derivation of (\ref{I_Lang_w_spectr})
we find:
\begin{equation}
\label{I_Lang_w_spectr_par}
  I_{\Omega,\omega}^{\|}=\frac{a_\nu k_0^{\nu-1} \left<E_L^2\right>  Q^4 |\cos\vartheta|}
  {\pi M^2c^4\gamma^6 }  \left(\frac{\omega}{\omega
  '}\right)^3 \times $$$$
  \left[1-\frac{\omega}{2\omega'\gamma_*^{2}}
\right]
   \left\{\left(\frac{\omega' -\omega\pe}{\bf nv}\right)^2
   +k_0^2\right\}^{-\nu/2}
   .
\end{equation}

Apparently, spectra (\ref{I_Lang_w_spectr}) and
(\ref{I_Lang_w_spectr_par}) look rather differently compared with
the spectrum (\ref{I_Lang_nw_2}) produced by a particle moving
transversely to vector ${\bf n}$. Given, in particular, that the
frequency $\omega'$ is directly linked with the emission angle
$\theta$, the $\delta$-function $\delta(\omega'-\omega\pe)$ permits
emitting a single frequency only in each direction. By comparison,
no $\delta$-function enters (\ref{I_Lang_w_spectr}) and
(\ref{I_Lang_w_spectr_par}), thus, a continuum spectrum rather than
distinct frequencies is emitted along any direction. Clearly, there
remains a distinct contribution to the emission intensity when
$\omega' \approx \omega\pe$. However, the range of the parameter
space where this resonant condition holds is relatively narrow, so
the "non-resonant" contribution from the remaining part of the
parameter space where $\omega' \neq \omega\pe$ can easily dominate
the resonant contribution. To see this explicitly, consider the
radiation intensity into the full solid angle by integration of
(\ref{I_Lang_w_spectr}) over $d\Omega=\sin\theta d\theta d\varphi
\approx 2\pi d(\omega'/\omega)$ that yields

\begin{equation}
\label{I_Lang_w_spectr_2}
  I_{\omega}^{\bot}=\frac{ a_\nu k_0^{\nu-1} \left<E_L^2\right>  Q^4 \sin^2\vartheta}
  {M^2c^4\gamma^2 |\cos\vartheta|} \int_{1/2\gamma_*^2}
  ^{\infty} d\left(\frac{\omega'}{\omega}\right)
  \left(\frac{\omega}{\omega
  '}\right)^2 \times $$$$
  \left[1-\frac{\omega}{\omega'\gamma_*^{2}} +
  \frac{\omega^2}{2\omega'^2\gamma_*^{4}}
  \right] %$$$$ %\\\\
   \left\{\left(\frac{\omega' -\omega\pe}{\bf nv}\right)^2
   +k_0^2\right\}^{-\nu/2}
   .
\end{equation}

%Although we cannot perform full analytical treatment of the
%spectrum,
Let us analyze essential properties of the DRL spectrum on the basis
of the asymptotic evaluation.  At low frequencies $\omega \ll
\omega\pe\gamma^2$, we can discard $\omega'$ in
(\ref{I_Lang_w_spectr_2}) everywhere in the braces except narrow
region of parameters when $\omega' \approx \omega\pe$. This means
that for $\omega \ll \omega\pe\gamma^2$ the integral is composed of
two contributions. The first of them, a non-resonant one, arises
from integration over the region, where $\omega' \ll \omega\pe$.
Here, the emission is beamed within the characteristic emission
angle of $\vartheta \sim \gamma^{-1}$ along the particle velocity.
The integral converges rapidly, and so it may be taken along the
infinite region, which produces a flat radiation spectrum, $I_\omega
\propto \omega^0$, or $I_\omega \propto \omega^2$ at lower
frequencies, $\omega < \omega\pe\gamma$. However, as far as
$\omega'$ approaches $\omega\pe$, a resonant contribution comes into
play. Now, in a narrow vicinity of $\omega\pe$, we can adopt
\begin{equation}
\label{del_Lang_approx}
     \left\{\left(\frac{\omega' -\omega\pe}{\bf nv}\right)^2
   +k_0^2\right\}^{-\nu/2} \propto \delta(\omega' -\omega\pe)
   ,
\end{equation}
which results in a single-wave-like contribution, $I_\omega \propto
\omega^1$. The full spectrum at $\omega < \omega\pe\gamma^2$,
therefore, is just a sum of these two contributions. At high
frequencies, $\omega \gg \omega\pe\gamma^2$, the term $\omega'$
dominates in the braces, so other terms can be discarded. Thus, a
power-law tail, $I_\omega \propto \omega^{-\nu}$ arises in this
spectral range.

\begin{figure}
%\keepaspectratio
%\vspace{1in}
\hspace{-0.4in}
\includegraphics[bb=18 5 282 217,height=2.5in]{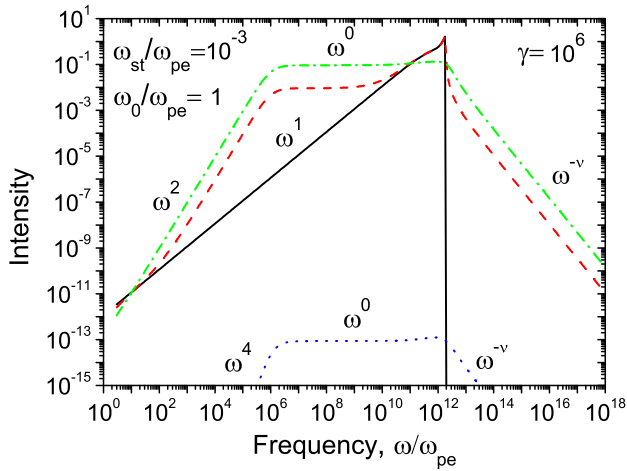}
\caption{\small \it DRL spectra produced by a particle with
$\gamma=10^6$ in a plasma with developed 1d Langmuir turbulence
for various particle propagation directions: $\cos\vartheta = 0$,
solid/black curve; $\cos\vartheta = 10^{-3}$, dashed/red curve;
$\cos\vartheta = 0.5$, dash-dotted/green curve; $\cos\vartheta =
1$, dotted/blue curve. Parameters are given in the Figure. The
"parallel" contribution (blue/dotted curve) is very small
($10^{-12}$) for the highly relativistic particle, although it
becomes competing for moderately relativistic particles. }
\label{DRL_1d}
\end{figure}

Although this integral cannot be taken analytically in a general
case,  it is easy to obtain corresponding spectra numerically.
Figure \ref{DRL_1d} displays a numerically integrated example of the
DRL spectra for various angles between ${\bf v}$ and ${\bf n}$ for
highly relativistic particle with $\gamma=10^6$. The black/solid
curve shows the DRL spectrum arising as the particle moves strictly
perpendicular to vector ${\bf n}$, which coincide with the spectrum
arising in the case of spatially uniform Langmuir oscillations
\cite{Tsytovich_Chikhachev_1969}. The spectrum consists of a rising
region $I_{\omega} \propto \omega^1$ at $\omega < 2
\omega\pe\gamma^2$ and drops abruptly to zero at $\omega > 2
\omega\pe\gamma^2$. Remarkably, there are prominent differences
between this spectrum and those generated for oblique particle
propagation even though the spectra can be similar to each other in
the immediate vicinity of the spectral peak.

To find the applicability region of the perturbation theory
applied above, we should estimate the characteristic deflection
angle of the emitting electron on the emission coherence length
$l_c=2c\gamma_*^{2}/\omega$, where the elementary emission pattern
is formed \cite{Fl_2005a}. Consider a simple source model
consisting of uncorrelated cells with the size $l_0=2\pi
c/\omega_0$, each of which contains coherent Langmuir oscillations
with the plasma frequency $\omega\pe$. Inside each cell the
electron velocity can change by the angle $\theta_0 \sim
\omega\st/(\omega\pe \gamma)$ if $\omega_0 \lsim \omega\pe$, where
$\omega_{st}=Q\left< E_{L}^2\right>^{1/2}/Mc$ (in the other case,
$\omega_0 > \omega\pe$, the results of \cite{Fl_2005a} apply).
Then, after traversing $N=l_c/l_0$ cells, the mean square of the
deflection angle is $\theta_c^2 =\theta_0^2 N \sim \omega\st^2
\omega_0/(\omega\omega\pe^2)$. The perturbation theory is only
applicable if this diffusive deflection angle is smaller than the
relativistic beaming angle, $\gamma^{-1}$, i.e., it is always
valid at sufficiently high frequencies $\omega
> \omega_* \equiv \omega\st^2 \omega_0 \gamma^2/\omega\pe^2$.
Note, that the bounding frequency $\omega_*$ increases with
$\omega_0$, while Diffusive Synchrotron Radiation (DSR) in random
magnetic field displays the opposite trend.  The perturbation theory
will be applicable to the entire DRL spectrum if the condition
$\theta_c^2 < \gamma^{-2}$ holds for the frequency $\omega\pe\gamma$
\cite{Fl_2005a}, where the coherence length of the emission has a
maximum. This happens for the particles whose Lorenz-factors obey
the inequality
\begin{equation}
 \label{DRL_criter}
\gamma < \omega\pe^3/(\omega\st^2\omega_0).
\end{equation}

Let's compare the DRL spectra calculated numerically (Figure
\ref{DRL_1d}) for the case of a broad turbulence distribution over
spatial scales with the DSR spectra in case of stochastic magnetic
fields \citep{Fl_2005a}. %If the Langmuir turbulence consists of
%relatively small-scale waves, $\omega_0 |\cos\vartheta| \equiv k_0 c
%|\cos\vartheta| \gsim \omega\pe$ then the shape of the spectrum is
%similar to the DSR spectrum because the variation of the electric
%field along the path of the relativistically moving particle will be
%mainly provided by spatial variations of the Langmuir turbulence
%rather than its temporal oscillations. However,
Apparently, there is a remarkable difference between these two
emission mechanisms, especially in the case of the long-wave
turbulence, $\omega_0 |\cos\vartheta| \equiv k_0 c |\cos\vartheta|
\ll \omega\pe$. First, we note that the perturbation theory of DRL
has a broader applicability region, in particular, it applies for
higher energy electrons than the perturbative version of the DSR
theory \citep{Fl_2005a} since the criterion (\ref{DRL_criter}) is
softer than the corresponding criterion in \cite{Fl_2005a}. This
happens because in the presence of the Langmuir turbulence the rapid
temporal oscillations of the electric field direction substantially
compensate angular deflections of the particle, so the average
trajectory is much more similar to a straight line than for the case
of the random magnetic field with the same $\omega_0$ and
$\omega\st$. Then, a distinct spectral peak at $\omega =
2\omega\pe\gamma^2$ is formed with the linear decrease of the
spectrum with frequency, which is not present in case of the DSR. At
lower frequencies, however, this falling part of the spectrum gives
way to a flat spectrum, which is entirely missing within the
one-wave approach \cite{Tsytovich_Chikhachev_1969} and absent for
exactly transverse particle motion. Position of the corresponding
turning point depends on the $\omega_0 \cos\vartheta /\omega\pe$
ratio in such a way that for $\omega_0 |\cos\vartheta| \gsim
\omega\pe$ the flat spectral region entirely dominates the range
from $\omega\pe\gamma$ to $\omega\pe\gamma^2$. It is worth
emphasizing that the deviations of the DRL spectrum from the
single-wave spectrum (read, from the transverse case,
$\cos\vartheta=0$) is prominent even for oblique propagation angles
only slightly different from $\pi/2$, e.g., $\cos\vartheta=10^{-3}$
as in Figure \ref{DRL_1d}. Therefore, the presence of a broad
turbulence spectrum considered here in detail results in important
qualitative change of the emission mechanism, which cannot generally
be reduced to a simplified treatment relying on the single-wave
approximation with some rms value of the Langmuir electric field.

%\section{Discussion}

%\newpage

Modern computer simulations of shock wave interactions, especially
in the relativistic case, suggest that the energy density of the
excited Langmuir turbulence can be very large
\cite{Dieckmann_2005}, e.g., far in excess the energy of the
initial regular magnetic field. In particular, at the shock wave
front the electric field can be as strong as the corresponding
wave-breaking limit, i.e., $\omega\st \sim \omega\pe$
\cite{Silva_2006}. In this case the random walk of relativistic
electrons in the stochastic electric field can give rise to
powerful contribution in the nonthermal emission of an
astrophysical object, entirely dominating full radiation spectrum
or some broad part of it.

Although any detailed application of the considered emission
process is beyond the scope of this letter, we mention that the
DRL is a promising mechanism for the gamma-ray bursts and
extragalactic jets. In particular, some of the prompt gamma-ray
burst emission displays rather hard low-energy spectra with the
\emph{photon} spectral index $\alpha$ up to 0. The DRL spectral
asymptote $I_\omega \propto \omega^{1}$, which appears just below
the spectral peak at $2\omega\pe\gamma^2$, fits well to those
spectra. Remarkably, the flat lower-frequency asymptote, $I_\omega
\propto \omega^{0}$, can account for the phenomenon of the X-ray
excess \citep{Preece_etal_1996,Sakamoto_etal_2005} and prompt
optical flashes accompanying some GRBs.

In addition, this mechanism along with the DSR in random magnetic
fields \citep{Fl_2005c} can be relevant to the UV-X-ray flattenings
observed in some extragalactic jets. For example
\cite{Jester_etal_2006},
%although full spatially resolved radio to X-ray spectra of the jet
%in M87 agrees well with the DSR model \citep{Fl_2005c}, for the
%jet in 3C~273 this agreement holds from the radio to UV range,
%while its X-ray emission seems to require an additional component.
%Alternatively,
X-ray observations of the jet in 3C~273 look inconsistent with the
standard synchrotron of DSR models. Remarkably, the entire
UV-to-X-ray spectrum of 3C~273 might be produced by DRL, which can
be much flatter than usual DSR   in the range $\omega\Be\gamma^2 \ll
\omega \ll \omega\pe\gamma^2$.

%\newpage
\vspace{0.2cm}

\section*{Acknowledgments}
\vspace{-0.2cm} This work was supported in part by the RFBR grants
06-02-16295, 06-02-16859, and 07-02-00245. We have made use of
NASA's Astrophysics Data System Abstract Service.

\vspace{-0.52cm}
%\end{acknowledgements}

%\bibliographystyle{mn2e} \bibliography{DSR_PWNs,DSR_Langmuir}
\bibliographystyle{apsrev} \bibliography{DSR_PWNs,DSR_Langmuir_PRL}

\end{document}